\documentclass[12pt]{iopart}
\usepackage{iopams}
\usepackage{epsfig}

\begin{document}

\tolerance = 10000


\title{The TTF finite-energy spectral features in photoemission of TTF-TCNQ:
The Hubbard-chain description}
\author{D. Bozi$^1$, J. M. P. Carmelo$^2$, K. Penc$^3$,
P. D. Sacramento$^4$} 
\address{$^1$Centro de F\'{\i}sica de Materiales, 
Centro Mixto CSIC-UPV/EHU,
E-20018 San Sebastian, Spain\\
$^2$GCEP-Center of Physics, U. Minho, Campus Gualtar,
P-4710-057 Braga, Portugal\\
$^3$Res. Inst. for Solid State Physics
and Optics, H-1525 Budapest, P.O.B. 49, Hungary\\
$^4$Departamento de F\'{\i}sica and CFIF, IST, 
Universidade T\'ecnica de Lisboa, 
Av. Rovisco Pais, 1049-001 Lisboa, Portugal\\
E-mail: carmelo@fisica.uminho.pt}

\begin{abstract}
A dynamical theory which accounts for all microscopic one-electron 
processes is used to study the spectral function of the 
1D Hubbard model for the whole $(k,\,\omega)$-plane, beyond 
previous studies which focused on the weight distribution in the 
vicinity of the singular branch lines only. While our predictions agree 
with those of the latter studies concerning the tetracyanoquinodimethane 
(TCNQ) related singular features in photoemission of the organic 
compound tetrathiafulvalene-tetracyanoquinodimethane 
(TTF-TCNQ) metallic phase, the generalized theory also 
leads to quantitative agreement concerning the tetrathiafulvalene  (TTF) 
related finite-energy spectral features, which are found to correspond 
to a value of the on-site repulsion $U$ larger than for TCNQ. 
Our study reveals the microscopic mechanisms behind the unusual 
spectral features of TTF-TCNQ and provides a good overall 
description of those features for the whole $(k,\,\omega)$-plane. 
\end{abstract}


\maketitle

Most early studies of quasi-one-dimensional (1D) conductors such as
tetrathiafulvalene-tetracyanoquinodimethane (TTF-TCNQ)
have focused on the various low-energy phases, which are
not metallic and correspond to broken-symmetry states \cite{Basista}. Recently, the
resolution of photoemission experiments has improved, and the {\it normal} state of these
compounds was found to display exotic spectral properties \cite{Zwick,Ralph,Claessen}. However,
such a metallic state refers to finite energies and thus is not described by the 
usual low-energy schemes. Based on a new quantum-object description
of the 1D Hubbard model \cite{Carmelo}, a preliminary version of the 
finite-energy dynamical theory considered here was used in the studies 
of Ref. \cite{Euro}, which provided the weight distribution in the vicinity 
of well-defined branch lines only. Those lines refer to theoretical singular 
spectral features, which were found to describe the sharpest TCNQ
related spectral dispersions observed in TTF-TCNQ by angle-resolved 
photoelectron spectroscopy (ARPES). Further details of such preliminary studies 
on the sharpest TCNQ related spectral features are provided 
in Ref. \cite{TCNQ-06}. In the mean while, a complete and more powerful 
version of the pseudofermion dynamical theory, which accounts for all
one-electron microscopic processes, was presented in Refs. \cite{V,ADD}. 
(Such a general finite-energy theory recovers the known low-energy
results in the limit of low energy, as confirmed in Ref. \cite{LE}.)  

In this paper the latter theory is suitably used to evaluate the momentum and energy 
dependence of the one-electron weight distribution of the model
for the whole $(k,\,\omega)$-plane. While both our predictions 
and those of Ref. \cite{Benthien} (which refer to the TCNQ features
only) agree with those of Ref. \cite{Euro} for the TCNQ related singular features, 
we are also able to derive a theoretical weight distribution for the TTF 
related spectral features. For TTF the best quantitative agreement between the 
theory and experiments is reached for values of $U$ larger than those 
preliminarily estimated in Ref. \cite{Euro} for the TTF stack of
molecules \cite{discrepancy}. The $U$ value 
found here for TTF is larger than for TCNQ, in agreement
with results from the low-energy broken-symmetry phase \cite{Basista}.
Our study clarifies the microscopic processes behind the unusual spectral 
properties of TTF-TCNQ and provides a good overall description of 
its spectral features for the whole $(k,\,\omega)$-plane. It also reveals 
that the electronic degrees of freedom of the {\it normal} state of
quasi-1D metals reorganize for all energies in terms of charge and spin
objects, whose scattering determines the unusual spectral properties. 

The structure of the quasi-1D conductor TTF-TCNQ consists of parallel linear
stacks of planar TTF and TCNQ molecules \cite{Basista,Zwick,Claessen}. Its partial charge
transfer is $0.59$ electrons from the donor (TTF) to the acceptor (TCNQ) and thus the
electronic densities are $n=1.41$ and $n=0.59$, respectively. Due to electronic
correlations, the optical properties of the metallic phase depart significantly from
Drude-free-electron behavior \cite{Basista} and the finite-energy electronic structure,
as probed by ARPES, deviates significantly from band-theory calculations 
\cite{Zwick,Ralph,Claessen}. (See Fig. 7 of Ref. \cite{Claessen}.) For energy values larger than the
transfer-integrals for electron inter-chain hopping, the 1D Hubbard model is
expected to provide a good description of such correlations in quasi-1D
conductors \cite{Claessen,Vescoli}. The model
describes $N=[N_{\uparrow}+N_{\downarrow}]$ spin-projection
$\sigma=\uparrow,\,\downarrow$ electrons with densities $n=N/N_a$ and
$m=[N_{\uparrow}-N_{\downarrow}]/N_a$ in an 1D lattice of $N_a$ sites.
Except in the momentum axis of Fig. 1, we use units of lattice constant one, so that
$0\leq n\leq2$. We denote the electronic charge by $-e$ and define the Fermi
momentum as $k_F=\pi\,n/2$ for $n<1$ (for electrons) and $k_F=\pi\,[2-n]/2$ for $n>1$
(for holes). The model includes a first-neighbor transfer-integral $t$, for electron
hopping along the chain, and an effective on-site Coulomb repulsion $U$. This is one of
the few realistic models for which one can exactly calculate all the energy eigenstates
and their energies \cite{Lieb,Carmelo}. Its low-energy spectrum belongs to the universal class
of the Tomonaga and Luttinger liquid (TTL) theory \cite{Schulz,Voit,LE}. 
However, its finite energy physics goes beyond (but is
related to) the TLL and, until recently, it was not possible to extract from the exact
solution the values of the matrix elements between the energy eigenstates. These are
needed for the study of the finite-energy one-electron spectral-weight distributions.
(The usual TLL theory can be considered as a special case of the finite-energy liquid
used here, as we summarize below.)

\begin{figure}
\includegraphics[scale=0.4]{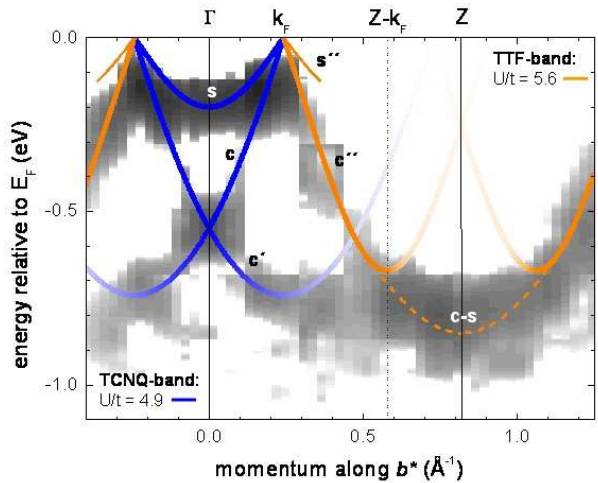} \caption{\label{fig1}
Experimental peak dispersions (grey scale) obtained by ARPES on TTF-TCNQ 
along the easy-transport axis as given in Fig. 7 of Ref. \cite{Claessen}
and matching theoretical branch and border lines. (The Z-point corresponds 
to the momentum $k=\pi$.) The corresponding detailed theoretical 
spectral-weight distributions over the whole $(k,\,\omega)$-plane are 
plotted below in Fig. 2. While the
theoretical charge-$c''$ and spin-$s''$ branch lines and $c-s$ 
border line refer to the TTF spectral features
found here ($n=1.41$, $U/t=5.61$), the charge-$c$, spin-$s$, and charge-$c'$ branch lines
correspond to the TCNQ dispersions ($n=0.59$, $U/t=4.90$) already studied
in Ref. \cite{Euro}.}
\end{figure}

Our study focuses on the theoretical description of the unusual spectral features 
associated with the TTF stacks, which until now has remained an open problem.
The model has both spin and $\eta$-spin $SU(2)$ symmetries. We denote by $\eta$ and
$\eta_z=-[N_a-N]/2$ (and $S$ and $S_z=-[N_{\uparrow}-N_{\downarrow}]/2$) the $\eta$-spin
value and projection (spin value and projection), respectively. For
$U/t\rightarrow\infty$ all energy eigenstates correspond to electronic occupancies with
fixed numbers of doubly occupied sites. However, the emergence of the exotic metallic
state involves an electron - rotated-electron unitary transformation, such that
rotated-electron double occupancy is a good quantum number for all $U/t$ values \cite{Carmelo}. 
As the Fermi-liquid quasiparticles, such rotated electrons have the same charge and spin as the
electrons, but refer to all energies and reorganize in terms of $[N_a-N_c]$ $\eta$-spin
$1/2$ holons, $N_c$ spin $1/2$ spinons, and $N_c$ spinless and $\eta$-spinless $c$
pseudoparticles, where $N_c$ is the number of rotated-electron singly occupied sites
\cite{Carmelo}. We use the notation $\pm 1/2$ holons and $\pm 1/2$ spinons, which refers
to the $\eta$-spin and spin projections, respectively. The $\pm 1/2$ holons of charge
$\pm 2e$ correspond to rotated-electron unoccupied $(+)$ and doubly-occupied $(-)$ 
sites. The complex behavior occurs for the $\sigma$-rotated
electrons occupying singly occupied sites: their spin degrees of freedom originate
chargeless $\sigma$ spinons, whereas their charge part gives rise to $\eta$-spinless and
spinless $c$ pseudoparticles of charge $-e$.
\begin{figure}
\includegraphics[scale=0.4]{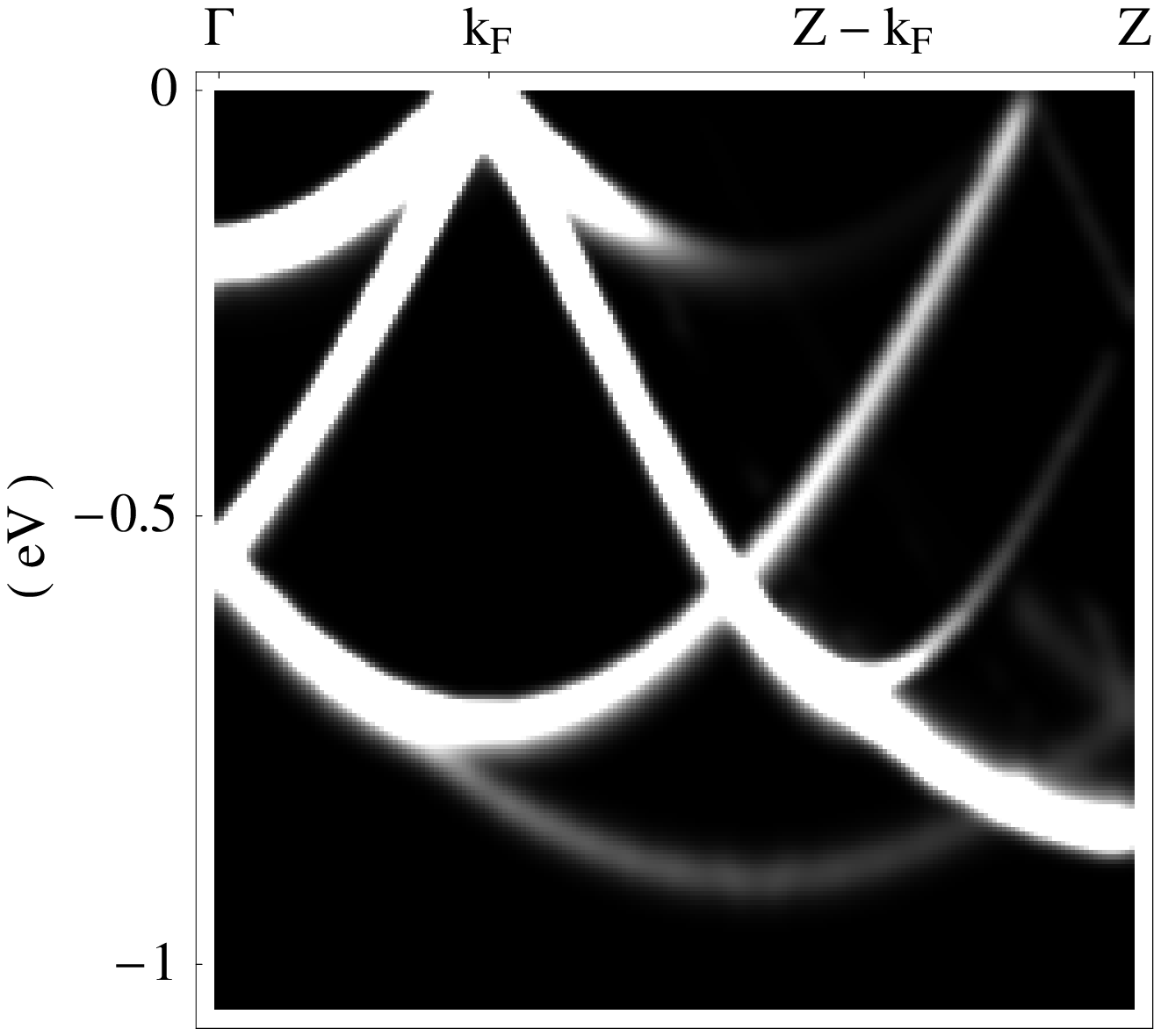}
\includegraphics[scale=0.4]{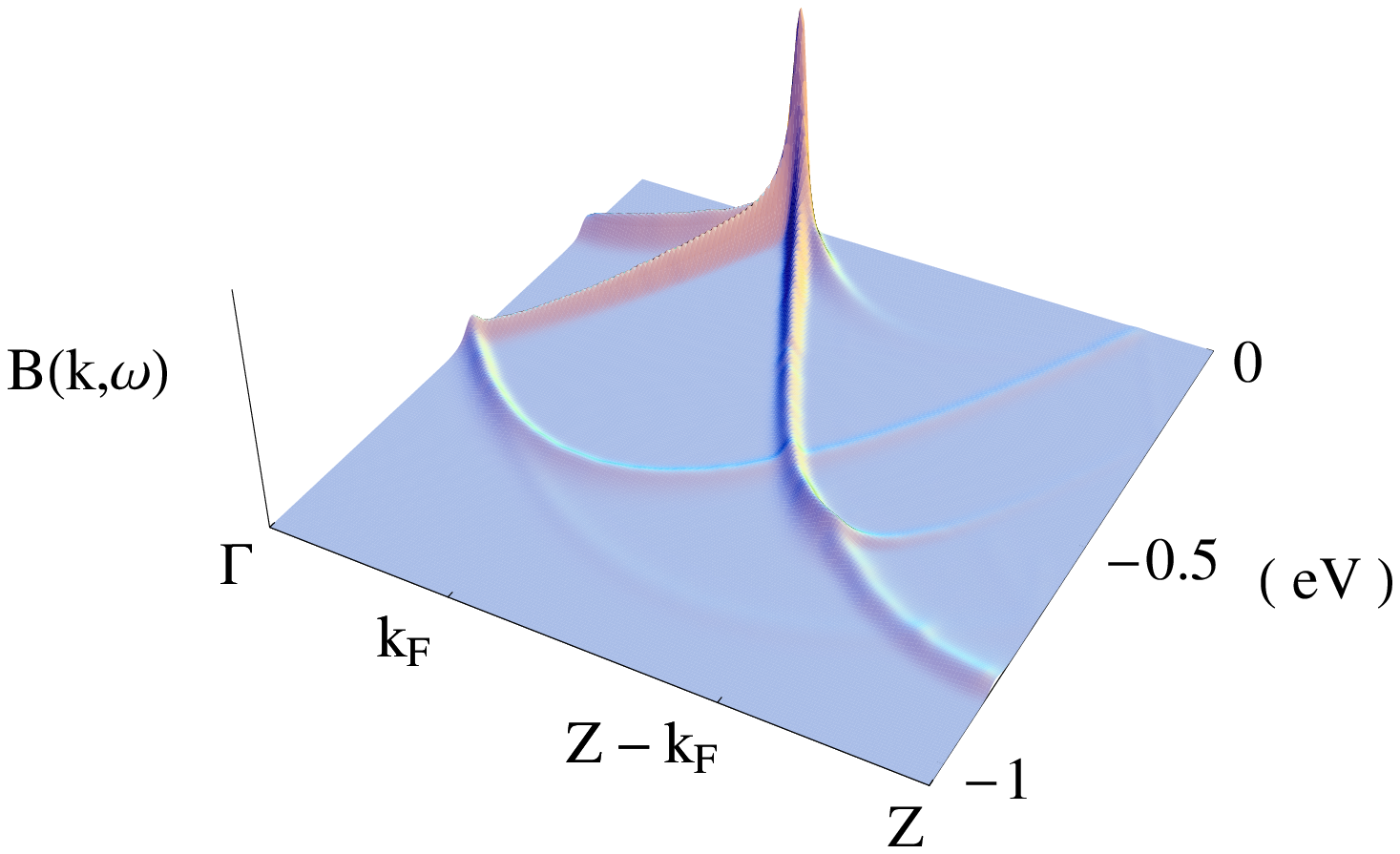}
\caption{\label{figure2} Full theoretical distribution of the one-electron 
removal spectral-weight intensity (left) and corresponding line shapes (right). 
The figures include both the TTF related spectral features for 
$n=1.41;t = 0.35\hspace{0.1cm}{\rm eV};U/t=5.61$ and 
those of TCNQ for $n=0.59;t=0.40\hspace{0.1cm}{\rm eV};U/t=4.90$, 
respectively, as in figure 1.}
\end{figure}

Based on symmetry considerations, we can classify the $\pm 1/2$ holons and $\pm 1/2$
spinons into two classes: those which remain invariant under the electron -
rotated-electron unitary transformation, and those which do not. The former are called
independent $\pm 1/2$ holons and independent $\pm 1/2$ spinons, with numbers reading
$L_{c,\,\pm 1/2}=[\eta\mp\eta_z]$ and $L_{s,\,\pm 1/2}=[S\mp S_z]$, respectively. The
latter are part of $\eta$-spin-zero $2\nu$-holon composite $c\nu$ pseudoparticles and
spin-zero $2\nu$-spinon composite $s\nu$ pseudoparticles, respectively, where
$\nu=1,2,...$ is the number of $+1/2$ and $-1/2$ holon or $+1/2$ and $-1/2$ spinon pairs.
The emergence of the exotic metallic state considered here involves a second unitary
transformation, which maps the $c$ pseudoparticles (and composite $\alpha\nu$
pseudoparticles) onto $c$ pseudofermions (and composite $\alpha\nu$ pseudofermions)
\cite{V}. Such a transformation introduces shifts of order $1/N_a$ in the pseudoparticle
discrete momentum values and leaves all other pseudoparticle properties invariant. As a
result of such momentum shifts and in contrast to the $c$ pseudoparticles and composite
$\alpha\nu$ pseudoparticles, the corresponding pseudofermions have no
residual-interaction energy terms \cite{V}.

As discussed below, the spectral-weight distribution of TTF-TCNQ is fully determined by
the occupancy configurations of the $c$ and $s1$ pseudofermions. We denote the latter by $s$
pseudofermions. These objects carry momentum $q$. For the $m=0$ ground-state there is $c$
pseudofermion occupancy for $\vert q\vert\leq 2k_F$ and unoccupancy for $2k_F<\vert
q\vert\leq \pi$, whereas the $s$ band is fully filled for $\vert q\vert\leq k_F$. (For
$m=0$, the exotic $s$ band has a momentum width of $2k_F$.) The momentum, $U/t$, and $n$
dependence of the $\alpha =c,\,s$ pseudofermion energy bands $\epsilon_{\alpha} (q)$,
group velocities $v_{\alpha}(q) =
\partial\epsilon_{\alpha}(q)/\partial q$,
and {\it Fermi-point} velocities $v_{c}\equiv v_{c}(2k_F)$ and $v_{s}\equiv v_{c}(k_F)$
is provided by the exact solution \cite{Carmelo,V}. Under the ground-state -
excited-state transitions, the ground-state $\alpha =c,\,s$ pseudofermions and holes are
scattered by the $\alpha' =c,\,s$ pseudofermions and holes created in these transitions.
Such zero-momentum-forward-scattering events lead to an overall phase shift
$Q_{\alpha}(q)/2=Q_{\alpha}^0/2+ Q^{\Phi}_{\alpha}(q,\{q'\})/2$, where $Q^{\Phi}_{\alpha}
(q,\{q'\}) =2\sum_{\alpha'=c,s}\, \sum_{q'}\,\pi\,\Phi_{\alpha,\,\alpha'}(q,q')\,\Delta
N_{\alpha'}(q')$,
$\pi\,\Phi_{\alpha,\,\alpha'}(q,q')=-\pi\,\Phi_{\alpha,\,\alpha'}(-q,-q')$ is a
two-pseudofermion phase shift whose $q$, $q'$, $n$, and $U/t$ dependence is provided by
the exact solution, the momentum-distribution deviation $\Delta N_{\alpha'}(q')$ is that
of the excited state, and $Q_{\alpha}^0/2=0,\pm\pi/2$ is a scattering-less phase shift
whose value is well defined for each transition. The excited-state-dependent {\it
Fermi-point} functionals $Q_{c}(\pm 2k_F)$ and $Q_{s}(\pm k_{F})$ fully control the
spectral properties \cite{V}. While the TLL theory involves a single
interaction-dependent spectral parameter, $1<\xi_0<\sqrt{2}$ \cite{Schulz,Voit,LE}, 
the description of the finite-energy spectral 
properties requires interaction and momentum-dependent phase
shifts $\pi\,\Phi_{c,\,c}(\pm 2k_F,q)$, $\pi\,\Phi_{c,\,s}(\pm 2k_F,q')$
$\pi\,\Phi_{s,\,c}(\pm k_{F},q)$, and $\pi\,\Phi_{s,\,s}(\pm k_{F},q')$, where
$2k_F<\vert q\vert<\pi$ for $n>1$, $0<\vert q\vert<2k_F$ for $n<1$, and $0<\vert
q'\vert<k_F$. In the limit of low-energy, the scattering centers are created in the
vicinity of the {\it Fermi-points} and our general description recovers the TLL theory
with $v_c(\pm 2k_F)=\pm v_c$, $v_s(\pm k_F)=\pm v_s$, and
$\xi_0=1+[\Phi_{c,\,c}(2k_F,2k_F)-\Phi_{c,\,c}(2k_F,-2k_F)]=
2\,[\Phi_{c,\,s}(2k_F,k_F)-\Phi_{c,\,s}(2k_F,-k_F)]$, as further discussed
in Ref. \cite{LE}.

A crucial test for the suitability of the model is whether the observed ARPES 
peak dispersions correspond to the theoretically predicted sharpest spectral 
features. In figure 1 we plot the positions of the sharpest theoretical spectral 
features considered below but omit the corresponding detailed spectral-weight distribution
over the $(k,\,\omega)$-plane predicted by the theory, which is plotted in Fig. 2. 
The figure also displays the experimental dispersions in the electron removal 
spectrum of TTF-TCNQ as measured by ARPES in Ref. \cite{Claessen}.
Figure 2 displays specifically the full theoretical distribution of the spectral-weight
intensity (left) and the line shapes (right) corresponding to the same values of $n$ and
$U/t$ as for the theoretical lines of Fig. 1. In the evaluation of the theoretical one-electron 
removal spectral features plotted in Figs. 1 and 2 we apply the improved pseudofermion 
dynamical theory of Ref. \cite{V}. One of
our goals is to find the value of $U/t$ for which the theoretical weight distribution
leads to the best agreement with the measured TTF related ARPES 
spectral features.

The total number of $\pm 1/2$ holons $(\alpha =c)$ and $\pm 1/2$ spinons $(\alpha =s)$
reads $M_{\alpha,\,\pm 1/2}=L_{\alpha,\,\pm 1/2}+\sum_{\nu
=1}^{\infty}\nu\,N_{\alpha\nu}$, where $N_{\alpha\nu}$ denotes the number of $\alpha\nu$
pseudoparticles. However, for the states which control the spectral properties of
TTF-TCNQ, one has that $N_{c\nu}=0$ for all $\nu$ and $N_{s\nu}=0$ for $\nu>1$ and for
$n<1$ (or $n>1$), $N_{c}=N$, $L_{c,\,+1/2}= [N_a-N]$, and $L_{c,\,-1/2}=0$ (or
$N_{c}=[2N_a-N]$, $L_{c,\,-1/2}= [N-N_a]$, and $L_{c,\,+1/2}=0$) and for $m>0$ (or
$m<0$), $N_{s1}=N_{\downarrow}$, $L_{s,\,+1/2}= [N_{\uparrow}-N_{\downarrow}]$, and
$L_{s,\,-1/2}=0$ (or $N_{s1}=N_{\uparrow}$, $L_{s,\,-1/2}=
[N_{\downarrow}-N_{\uparrow}]$, and $L_{s,\,+1/2}=0$). Since the independent holons and
spinons of these states are scatter-less objects \cite{V}, the weight distribution is
fully determined by the occupancy configurations of the $c$ and $s\equiv s1$
pseudofermions. For the $(k,\,\omega)$-plane regions of the ARPES data, the method used
in our calculation involves specific processes associated with ground-state -
excited-state transitions. For those which generate the dominant contributions
one $s$ hole is created and for densities $n>1$ and $n<1$
a $c$ pseudofermion and a $c$ hole, respectively, is created. The low-energy TLL
corresponds to processes where both such objects are created at momentum values in the
vicinity of their {\it Fermi points}. Since the low-energy phase of TTF-TCNQ is not
metallic and corresponds to broken-symmetry states, our results are to be applied for
processes with energies larger than the gap, beyond the reach of TLL theory.

For finite energy and $U/t$ values all sharp spectral features are of power-law type,
controlled by negative exponents. Important finite-energy processes are those where one
$\alpha$ pseudofermion or hole is created at $q$ away from the {\it Fermi points} and the
second object is created at one of these points. The preliminary studies of Ref. \cite{Euro}
only considered such processes. They originate features centered on lines, $\omega
=\omega_{\alpha}(q)=\pm\epsilon_{\alpha}(q)$, in the $(k,\,\omega)$-plane. In the
vicinity and just below these lines, the spectral-function reads \cite{V,ADD},
\begin{equation}
B (k,\,\omega)\approx C_{\alpha} (q)\,(\omega_{\alpha}(q)-\omega)^{\zeta_{\alpha} (q)} \,
. \label{B-bl}
\end{equation}
When $\zeta_{\alpha} (q)<0$, the spectral feature is a singular branch line. The exponent
reads $\zeta_{\alpha}=-1+\zeta_0 (q)$, where $\zeta_0 (q)$ is a functional whose values
are fully controlled by the pseudofermion scattering. It reads, $\zeta_0 (q)
=\sum_{\iota=\pm 1}\sum_{\alpha =c,\,s}2\Delta_{\alpha}^{\iota} (q)$ where
$2\Delta_{\alpha}^{\iota}(q)\equiv (\iota\,\Delta N_{\alpha,\,\iota}^F+ Q^{\Phi}_{\alpha}
(\iota\,q^0_{F\alpha},q)/2\pi)^2$, $Q^{\Phi}_{\alpha} (\iota\,q^0_{F\alpha},q)/2$ is the
scattering part of the overall phase shift defined above, $\iota\,\Delta
N^F_{\alpha,\,\iota}=\Delta q_{F\alpha,\,\iota}/[2\pi/N_a]$, and $\Delta
q_{F\alpha,\,\iota}$ with $\iota =\pm 1$ is the {\it Fermi point} deviation relative to
the ground-state values $\iota\,q^0_{Fc}=\iota\,2k_F$ or $\iota\,q^0_{Fs}=\iota\,k_F$.
Expression (\ref{B-bl}) does not apply in the TLL regime, which corresponds to $k\approx
\pm k_F$ in Fig. 1, where the power-law exponent has a different
expression \cite{V}. The two regimes are separated by a small crossover
region. Thus, the finite-energy normal state found here for TTF-TCNQ 
{\it cannot} be described by the usual TLL.

There are other important types of finite-energy processes which were not considered
in the preliminary studies of Ref. \cite{Euro}. Those which involve creation of
more than two quantum objects lead to very little weight. In turn, some of the
processes where both a $c$ pseudofermion or hole and a $s$ hole are created 
at momentum values $q$ and $q'$, respectively, away from the {\it Fermi points} 
are important. When $v_{c}(q) \neq v_{s}(q')$, such processes do not lead
to singular spectral features and generate the background spectral weight 
all over the $(k,\,\omega)$-plane, which although being in general small must 
be accounted for. (See Fig. 2.) Furthermore, a second type of sharp feature 
not considered in Ref. \cite{Euro} corresponds to lines generated by such
processes when both created objects move with the same group velocity, $v_{c}(q)
=v_{s}(q')$, and the spectral feature corresponds to a border line, $\omega =\omega_{BL}
(k)=[\pm\epsilon_{c}(q)-\epsilon_{s}(q')]\,\delta_{v_{c}(q) ,\,v_{s}(q')}$, in the
$(k,\,\omega)$-plane. The spectral function reads \cite{ADD},
\begin{equation}
B (k,\,\omega)\approx C_{BL} 
(k)\,(\omega -\omega_{BL} (k))^{-{1/2}}\,
, \label{B-bol}
\end{equation}
in the vicinity and just above such a line. The TTF line called $c-s$ in Fig. 1 and the
weaker TCNQ bottom line of Fig. 2 are of this type. The latter weaker theoretical
line is not plotted in Fig. 1, yet it clearly marks the lower limit of the experimental
weight distribution. In the limit of $U/t>>1$, our one-electron weight distributions
agree with those of Ref. \cite{Penc}.

By careful analysis of the $k$, $\omega$, and $U/t$ dependence of the obtained
theoretical weight distribution, we find that for $n=1.41$ the electron removal spectrum
calculated for $t=0.35$ eV and $U=1.96$ eV ($U/t=5.61$) yields the best agreement with
the TTF experimental dispersions. (The $U/t=5.61$ TTF value is much larger than that 
preliminarily estimated in Ref. \cite{Euro}.) Remarkably, the only fitting parameter is $U/t$. For
the considered values of $n$ and $U/t$, the singular charge-$c''$ and spin-$s''$ branch
lines and the singular $c-s$ border line of Fig. 1 correspond to the sharp spectral features of
the model one-electron removal spectral function for the ARPES $(k,\,\omega)$-plane
region. The fading parts of the theoretical charge-$\alpha =c',\,c''$ branch lines of
Fig. 1, not seen in the experiment, correspond to values of the momentum where the
constant $C_{\alpha} (q)$ of the expression (\ref{B-bl}) in the vicinity of these lines
is small. Although there is a reasonably good overall quantitative agreement between
the theoretical one-electron weight distributions and the TTF related features measured 
by ARPES, there are apparent differences in the finest details, for instance the 
broadening of some of the sharp features predicted by the theory. However and in 
spite of the recent improvements in the resolution of 
photoemission experiments \cite{Zwick,Ralph,Claessen}, it is difficult to measure 
the weight-distribution finest details experimentally, in part due to the extrinsic losses that occur on 
anisotropic conducting solids \cite{Joynt}. Hence while our theoretical
description provides the dominant microscopic processes
behind the overall unusual spectral-weight distribution observed
in the real material, it is difficult to judge which other smaller effects 
may play some role in the weight-distribution finest details. 

Our general study refers to the whole $(k,\,\omega)$-plane and confirms 
the validity of the predictions of Ref. \cite{Euro} for
the TCNQ related spectral features: for $n=0.59$ the finite-energy-electron-removal spectrum
calculated for $t = 0.40$ eV and $U=1.96$ eV ($U/t=4.90$) yields an almost perfect
agreement with the TCNQ experimental dispersions, which correspond to the spin-$s$,
charge-$c$, and charge-$c'$ branch lines of Fig. 1 and the weaker border line shown in
Fig. 2. There we plot the full theoretical distribution of the weight intensity
resulting from electron removal both for $n=1.41;U/t=5.61$ and $n=0.59; U/t=4.90$ and the
corresponding line shapes, respectively.

Our results reveal that the ARPES peaks refer to separate spin ($s$ hole) and
charge ($c$ pseudofermion or hole) objects for the whole energy bandwidth, whose
line-shape depends on the interaction. For the Hubbard model, such a
spin-charge separation persists in the limit of low energy, where the quantum liquid
becomes a TLL. An important exception is the TTF singular border line named $c-s$ in Fig.
1 (and the weaker TCNQ border line shown in Fig. 2), which refers to a charge and
a spin object moving with the same velocity. Before merging this line, the charge-$c''$ branch line
refers to $q$ values such that $\vert v_{c}(q)\vert > v_s$, whilst for $q$ and $q'$
obeying the relation $v_{c} (q)=v_{s} (q')$ such that $0\leq \vert v_{c} (q)\vert = \vert
v_{s} (q')\vert\leq v_s$ the singular $c-s$ border line emerges. Such a feature does not
exist at low energy because of differing charge and spin velocities.

The transfer-integral values obtained here are about twice as large as those found by
band theory, consistently with the experimental bandwidth being much larger than
predicted by traditional estimates \cite{Zwick}. Moreover, our values for $U/4t$ are of
the order of unity and larger for TTF ($U/t=5.61$) than for TCNQ ($U/t=4.90$),
consistently with the TTF-TCNQ broken-symmetry-states and optical properties
\cite{Basista}. The effects of the onsite repulsion $U$ considered here for 
the electrons inside the solid, lead to a weight distribution that agrees with the general
ARPES spectrum structure over a wide range of finite energies. (This is in contrast
to band-theory calculations, as confirmed in Fig. 7 of Ref. \cite{Claessen}.) However, 
when the photoelectron is in the vacuum above the crystal it may create excitations 
in the substrate via long-ranged interactions beyond our model. The resulting 
inelastic losses as well other effects of finite temperature and long-range
Coulomb interactions \cite{Assaad} are expected to be the mechanisms behind 
the broadening of the singular features predicted here \cite{Joynt}. Such
effects lead to the broad peaks observed in the ARPES of Ref. \cite{Claessen} 
but do not change their overall distribution over the $(k,\omega)$ plane, which 
remains as found in this paper.

We thank R. Claessen, N. M. R. Peres, T. C. Ribeiro, and 
M. Sing for discussions and the support of ESF Science Program INSTANS, 
European Union Contract 12881 (NEST), FCT grants 
SFRH/BD/6930/2001, POCTI/FIS/58133/2004, and PTDC/FIS/64926/2006
and OTKA grant T049607.


\begin{thebibliography}{99}
\bibitem{Basista}
	Basista H, Bonn D A, Timusk T, Voit, J\'erome D
	and Bechgaard K 1990
        \emph{Phys. Rev. B}{\bf 42} 4088.
\bibitem{Zwick}
	Zwick F, J\'erome D, Margaritondo G, 
	Onellion M, Voit J and Grioni M 1998
	\emph{Phys. Rev. Lett.}{\bf 81} 2974.
\bibitem{Ralph} 
	Claessen R, Sing M, Schwingenschl\"ogl U, Blaha P,
        Dressel M and Jacobsen CS 2002
        \emph{Phys. Rev. Lett.} {\bf 88} 096402.
\bibitem{Claessen}
	Sing M, Schwingenschl\"ogl U, Claessen R,
        Blaha P, Carmelo JMP, Martelo LM, Sacramento PD,
        Dressel M and Jacobsen CS 2003
        \emph{Phys. Rev. B}{\bf 68} 125111.
\bibitem{Carmelo}
	Carmelo JMP, Rom\'an JM and Penc K 2004
        \emph{Nucl. Phys. B}{\bf 683} 387.
\bibitem{Euro}
	Carmelo JMP, Penc K, Martelo LM, Sacramento PD, 
	Lopes dos Santos JMB, Claessen R, Sing M and Schwingenschl\"ogl U
	2004 \emph{Europhys. Lett.}{\bf 67} 233.
\bibitem{TCNQ-06}	
	Carmelo JMP, Penc K, Sacramento PD, Sing M and 
	Claessen R 2006 \emph{J. Phys.: Cond. Matt.}{\bf 18} 5191.
\bibitem{V}
        Carmelo JMP, Penc K and Bozi D 2005
        \emph{Nucl. Phys. B}{\bf 725} 421; 2006
        \emph{Nucl. Phys. B (erratum)}{\bf 737} 351;  
        Carmelo JMP and Penc K 2006
        \emph{Eur. Phys. J. B}{\bf 51} 477.
        Carmelo JMP and Penc K 2006
        \emph{J. Phys.: Cond. Mat.}{\bf 18} 2881.
\bibitem{ADD}
        Carmelo JMP and Bozi D 2007, submitted to Annals of Physics.		
\bibitem{LE}         
        Carmelo JMP, Martelo LM and Penc K 2006 
        \emph{Nucl. Phys. B}{\bf 737} 237; 
	Carmelo JMP and Penc K 2006
	\emph{Phys. Rev. B}{\bf 73} 113112.
\bibitem{Benthien}
	Benthien H, Gebhard F and Jeckelmann E 2004
	\emph{Phys. Rev. Lett.}{\bf 92} 256401.
\bibitem{discrepancy} 
	The reason for such a discrepancy is that the present method takes into
	account all spectral features distributed over the whole $(k,\,\omega)$-plane, 
	whereas the analysis of Ref. \cite{Euro} relied on the momentum and
	energy dependence in the vicinity of the branch lines only.
\bibitem{Vescoli}
	Vescoli V, Degiorgi L, Henderson W, Gr\"uner C, Starkey KP
	and Montgomery LK 1998
	\emph{Science}{\bf 281} 181.
\bibitem{Lieb}
        Lieb Elliot H and Wu FY 1968
        \emph{Phys. Rev. Lett.}{\bf 20} 1445;        
        Martins MJ and Ramos PB 1998
        \emph{Nucl. Phys. B}{\bf 522} 413.
\bibitem{Schulz}
        Schulz HJ 1990
        \emph{Phys. Rev. Lett.}{\bf 64} 2831.
\bibitem{Voit}
	Voit J 1995
        \emph{Rep. Prog. Phys.}{\bf 58} 977.         
\bibitem{Penc}
        Penc K, Hallberg K, Mila F and Shiba H 1996
        \emph{Phys. Rev. Lett.}{\bf 77} 1390.
\bibitem{Joynt}
        Joynt R 1999
        \emph{Science}{\bf 284} 777;
        Mills DL 2000
        \emph{Phys. Rev. B}{\bf 62} 11197; 
        Joynt R 2002
        \emph{Phys. Rev. B}{\bf 65} 077403.
\bibitem{Assaad}        
        Abendschein A and Assaad FF 2006
	\emph{Phys. Rev. B}{\bf 73} 165119;
        Bulut N, Matsueda H, Tohyama T and Maekawa S 2006 
        \emph{Phys. Rev. B}{\bf 74} 113106.        
\end{thebibliography}
\end{document}